\documentclass[conference,a4paper]{IEEEtran}
\IEEEoverridecommandlockouts
\usepackage{cite}
\usepackage{amsmath,amssymb,amsfonts}
\usepackage{algorithmic}
\usepackage{graphicx}
\usepackage{textcomp}
\usepackage{booktabs}
\usepackage{xcolor}
\usepackage{url}
\usepackage{multirow}

\def\BibTeX{{\rm B\kern-.05em{\sc i\kern-.025em b}\kern-.08em
    T\kern-.1667em\lower.7ex\hbox{E}\kern-.125emX}}
\begin{document}

\title{ECG classification using Deep CNN and Gramian Angular Field}

\author{\IEEEauthorblockN{Youssef Elmir}
\IEEEauthorblockA{\textit{Laboratoire LITAN} \\
\textit{École supérieure en Sciences et Technologies}\\
\textit{de l’Informatique et du Numérique}\\
RN 75, Amizour 06300, Béjaia, Algérie \\
SGRE-lab, University of Béchar, Algérie\\
elmir@estin.dz}
\and
\IEEEauthorblockN{Yassine Himeur}
\IEEEauthorblockA{\textit{College of Engineering and IT} \\
\textit{University of Dubai}\\
Dubai, UAE \\
yhimeur@ud.ae}
\and
\IEEEauthorblockN{Abbes Amira}
\IEEEauthorblockA{\textit{Department of Computer Science} \\
\textit{University of Sharjah}\\
Sharjah, UAE \\
aamira@sharjah.ac.ae}

}

\maketitle

\begin{abstract} 
This paper study provides a novel contribution to the field of signal processing and DL for ECG signal analysis by introducing a new feature representation method for ECG signals. The proposed method is based on transforming time frequency 1D vectors into 2D images using Gramian Angular Field transform. Moving on, the classification of the transformed ECG signals is performed using Convolutional Neural Networks (CNN). The obtained results show a classification accuracy of 97.47\% and 98.65\% for anomaly detection. Accordingly, in addition to improving the classification performance compared to the state-of-the-art, the feature representation helps identify and visualize temporal patterns in the ECG signal, such as changes in heart rate, rhythm, and morphology, which may not be apparent in the original signal. This has significant implications in the diagnosis and treatment of cardiovascular diseases and detection of anomalies.
\end{abstract}


\section{Introduction}
\subsection{Background}
Remote monitoring devices, whether worn or implanted, have revolutionized healthcare for patients with periodic heart arrhythmia by providing constant monitoring of heart activity. However, these devices generate large amounts of electrocardiogram (ECG) data that require interpretation, adding to the already heavy workload of physicians, nurses, and other medical staff. This increased workload contributes to the fatigue experienced by medical professionals \cite{weimann2021transfer}, leading to a higher risk of medical errors. Moreover, due to their high sensitivity to abnormalities, remote monitoring devices often trigger false alarms, resulting in a significant number of unnecessary ECG recordings. As a result, there is a growing demand for assistance in the interpretation of ECG recordings to support physicians in their clinical decision-making \cite{liu2021deep}.

Globally, millions of ECG recordings are collected annually, with the majority being analyzed and interpreted by computer algorithms. This necessitates ECG interpretation methods to be not only fast and accurate, but also applicable to various patients and devices \cite{chen2022automated,huang2023novel}. With the increasing digitization of ECG data and advancements in deep learning (DL) techniques, which can handle large amounts of raw data, there are new opportunities to enhance automated ECG interpretation. Recently, deep neural networks (DNNs) have demonstrated performance comparable to cardiologists when trained on a large dataset (n = 91,232) of raw ECG recordings \cite{wang2023hierarchical,xia2023generative}.

However, available ECG datasets can be much smaller, especially due to privacy and security issues, posing challenges in achieving desired performance levels \cite{mohebbian2023semi}. 
Additionally, ECG classification faces several challenges related to feature representation, including signal variability, noise and artifacts, dimensionality, interpretability, and scalability \cite{wang2023hierarchical}. Typically, ECG signals can vary significantly from patient to patient and can be affected by various factors such as age, gender, medication, and underlying medical conditions. This variability makes it challenging to identify relevant features that can be used for classification \cite{zheng2023phonocardiogram}. Moving on, ECG signals can be affected by various types of noise and artifacts, such as motion artifacts, baseline wander, and muscle noise. These artifacts can interfere with the accurate extraction of features and can lead to misclassification. Moving on, ECG signals are high-dimensional, and the feature space can be very large, making it difficult to identify the most informative features \cite{kumar2023investigation}.
Most importantly, some feature representation methods, such as DL techniques, can produce highly abstract and non-interpretable features, making it challenging to understand the underlying mechanisms of ECG classification. Moreover, some feature representation methods may not be scalable to large datasets, making it challenging to apply them to real-world applications \cite{holgado2023characterization}.
On the other hand, training deep neural networks (DNNs) to classify ECG data from remote monitoring devices presents several challenges \cite{copiaco2023innovative,sayed2023time}. These challenges include a severe class imbalance due to the infrequent occurrence of certain cardiovascular events, poor signal quality characterized by low sampling frequency, single ECG lead, and noise, as well as limited annotations due to the cost of manual labeling by experts \cite{kent2023fourier}.

\subsection{Contributions}
In this study, we first transform ECG signals into 2D representations using GAF. Moving froward, obtained images are then fed into different CNN models with various layers, including CNN (with eight layers), VGG-16, ResNet and EfficientNet to extract robust features from the input GAF images.
The proposed scheme has been assessed on a publicly available dataset, which includes two sets of heartbeat signals that are obtained from two well-known datasets related to heartbeat classification. The first one refers to the MIT-BIH Arrhythmia Dataset\footnote{\url{https://www.physionet.org/physiobank/database/mitdb/}} and the second one is the PTB Diagnostic ECG Database\footnote{\url{https://www.physionet.org/physiobank/database/ptbdb/}}.
The proposed method has enabled us to achieve an impressive accuracy and F1-score of 98.65\% using the EfficientNet model. These results outperform recently reported findings in classifying similar types of arrhythmias.

\section{METHODOLOGY}
\begin{figure}[ht]
\centering
\includegraphics[width=0.3\textwidth]{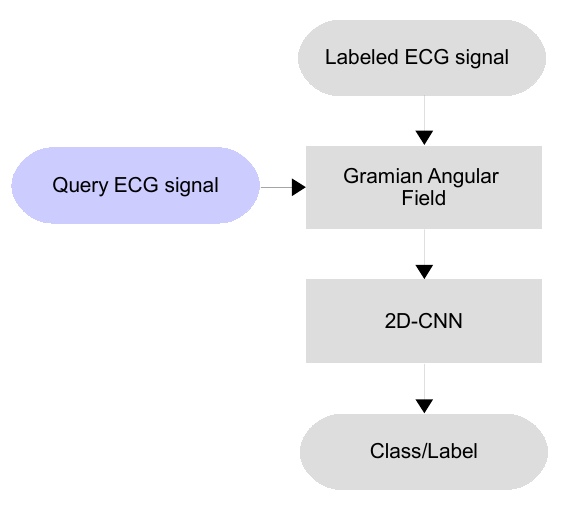}
\caption{Architecture of the proposed ECG representation and classification system.}
\label{fig:methodology}
\end{figure}

Figure \ref{fig:methodology} illustrates the architecture of the proposed ECG representation and classification system.
\label{sec:methodology}

\subsection{Feature representation using Gramian Angular Field}

\begin{figure}[ht]
\centering
\includegraphics[width=0.5\linewidth]{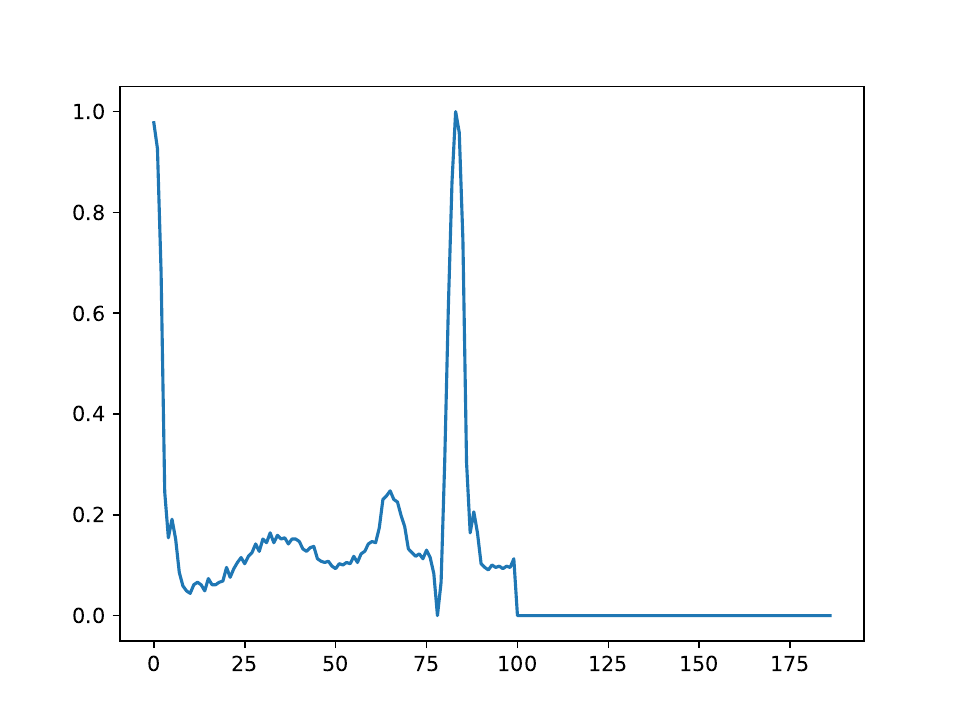}
\\(a) 1D ECG vector \\
\includegraphics[width=0.7\linewidth]{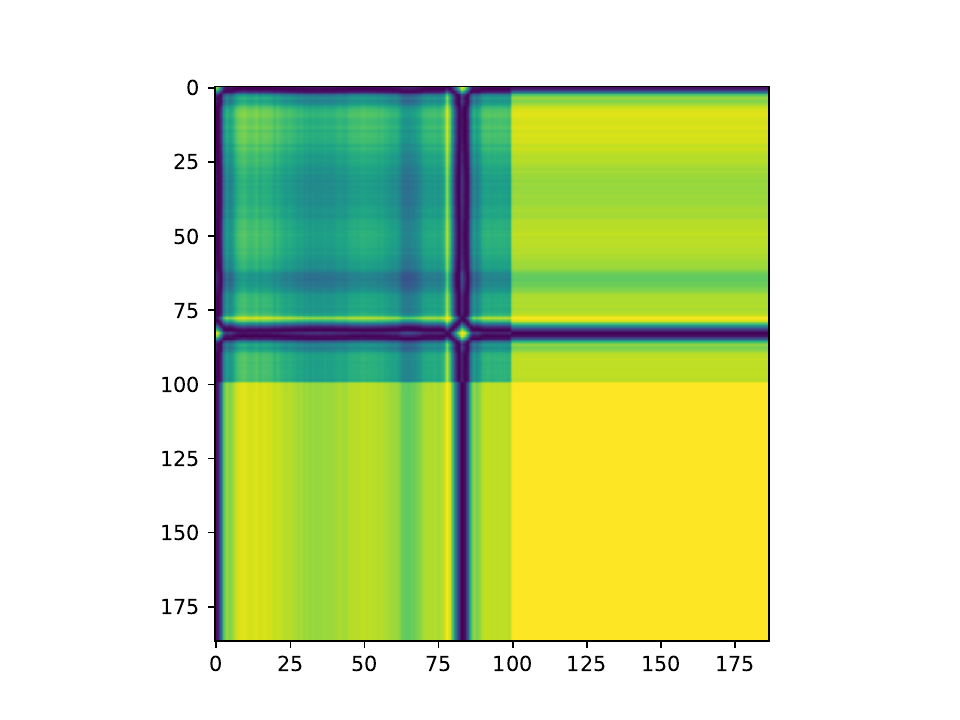}
\\(b) ECG 2D GAF image
\caption{Example of transforming a 1D ECG signal (of a sample) to a 2D GAF image: (a) 1D vector, and (b) ECG 2D GAF image.}
\label{fig:2}
\end{figure}

This study employs a Python library that was developed and described in a previous work by Wang et al\cite{wang2015encoding}. in 2015. The library is used to transform ECG signals into images. ECG signals are represented in a polar coordinate system instead of the typical Cartesian coordinates.
In the Gramian matrix, each element is actually the cosine of the summation of angles. Inspired by previous work on the
duality between time series and complex networks \cite{campanharo2011duality}. Figure \ref{fig:2} portrays an example of transforming a 1D ECG signal from the training datase to a 2D GAF image.
The size of the obtained images depends on the size of the length of the 1D vector. The output of GAF is a 2D squared image that has width and height equals to the length of the original 1D vector. In our case the length of the 1D vector was 187 and the size of the obtained image is 187x187. 

Regarding the big number of samples in the used dataset (87554 sample), size reduction is applied to all obtained GAF images, by consequence their unified size became 32x32.

\subsection{CNN model}
\begin{figure}[!t]
\centering
\includegraphics[width=1\linewidth]{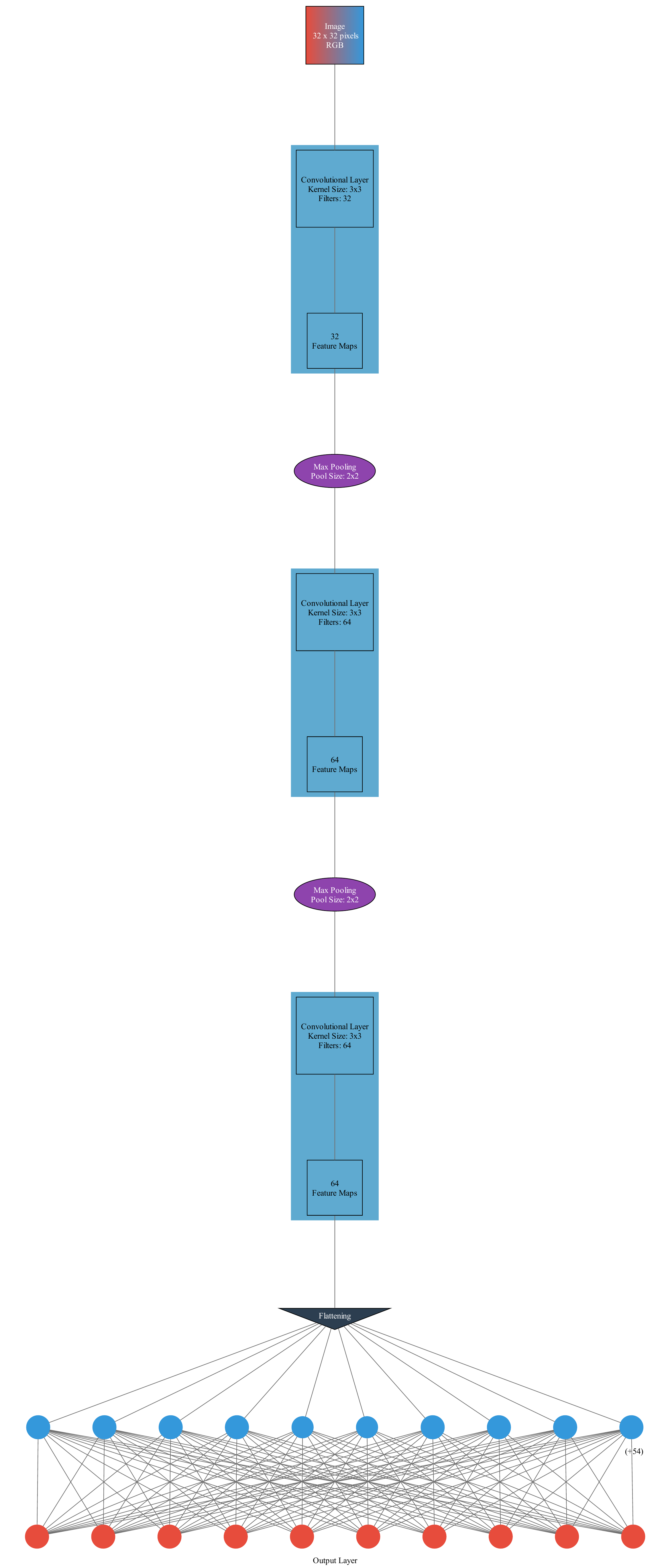}
\caption{Architecture of the used CNN for classification task}
\label{fig:imagearch}
\end{figure}

The architecture of the used CNN for ECG classification is presented in Figure \ref{fig:imagearch}. The proposed model consists of several layers, including convolutional layers, max pooling layers, and dense (fully connected) layers. The overall architecture of the model can be summarized as follows:
\begin{enumerate} 
\item Conv2D layer with 32 filters, a kernel size of (3, 3), 'relu' activation function, and an input shape of (32, 32, 3). This layer performs convolutional operation on the input image data.

\item MaxPooling2D layer with a pool size of (2, 2). This layer performs max pooling operation to downsample the feature maps obtained from the previous convolutional layer.

\item Conv2D layer with 64 filters and a kernel size of (3, 3) with 'relu' activation function. This layer performs another round of convolutional operation on the downsampled feature maps.

\item MaxPooling2D layer with a pool size of (2, 2). This layer performs another max pooling operation to further downsample the feature maps.

\item Conv2D layer with 64 filters and a kernel size of (3, 3) with 'relu' activation function. This layer performs one more round of convolutional operation.

\item Flatten layer, which flattens the 2D feature maps into a 1D vector, preparing it for the fully connected layers.

\item Dense layer with 64 units and 'relu' activation function. This layer serves as a fully connected layer, performing computation on the flattened feature vector.

\item Dense layer with 10 units, which represents the output layer. This layer does not have an activation function, as it is used for multi-class classification and will output raw scores for each class.
\end{enumerate} 

For comparison purposes, the same dense and output layers were used to define some of predesigned and pretrained CNNs (VGG16, ResNet50, and EfficientNet) in the experiments. These pretrained models have been previously trained on large image datasets, and their architectures have shown effectiveness in various computer vision tasks. 

\section{RESULTS \& DISCUSSION}


The dataset includes two sets of heartbeat signals that are obtained from two well-known datasets related to heartbeat classification, which are the MIT-BIH Arrhythmia Dataset\footnote{\url{https://www.physionet.org/physiobank/database/mitdb/}} and The PTB Diagnostic ECG Database\footnote{\url{https://www.physionet.org/physiobank/database/ptbdb/}}. The dataset is sufficiently large for training a deep neural network. It has been used to investigate the use of deep neural network architectures in heartbeat classification and to explore transfer learning capabilities. The signals in the dataset represent ECG shapes of heartbeats, including normal and abnormal cases affected by various arrhythmias and myocardial infarction. The signals are preprocessed and segmented into individual heartbeats.

\begin{table*}[!t]
\renewcommand{\arraystretch}{1.2}
\caption{Comparison of results on two ECG classification datasets}
\begin{center}
\begin{tabular}{c|c|c|c|c|c}
\hline
\textbf{Dataset} & \textbf{\textit{Number of Samples (train, test)}}& \textbf{\textit{Number of Categories (classes)}}& \textbf{\textit{Method}}& \textbf{\textit{Accuracy (\%)}}& \textbf{\textit{F1-score (\%)}} \\
\hline
\multirow{4}{*}{Arrhythmia} & \multirow{4}{*}{109446 (87554/21892)} & \multirow{4}{*}{5}    & CNN & 97.47 & 97.34 \\\cline{4-6}
                        & & & VGG16 & 97.32 & 97.18 \\\cline{4-6}
                        & & & ResNet50 & 97.07 & 96.92 \\\cline{4-6}
                        & & & EfficientNet & 96.78 & 96.64 \\\hline

\multirow{4}{*}{The PTB Diagnostic} & \multirow{4}{*}{14552 (11642/2910)} & \multirow{4}{*}{2} & CNN & 97.56 & 97.56 \\\cline{4-6}
                                    & & & VGG16 & 89.17 & 88.98 \\\cline{4-6}
                                    & & & ResNet50 & 95.84 & 95.90 \\\cline{4-6}
                                    & & & EfficientNet & 98.65 & 98.65 \\\hline

\end{tabular}
\label{tab1}
\end{center}
\end{table*}









\begin{figure*}[!t]
\centering
\includegraphics[width=0.435\linewidth]{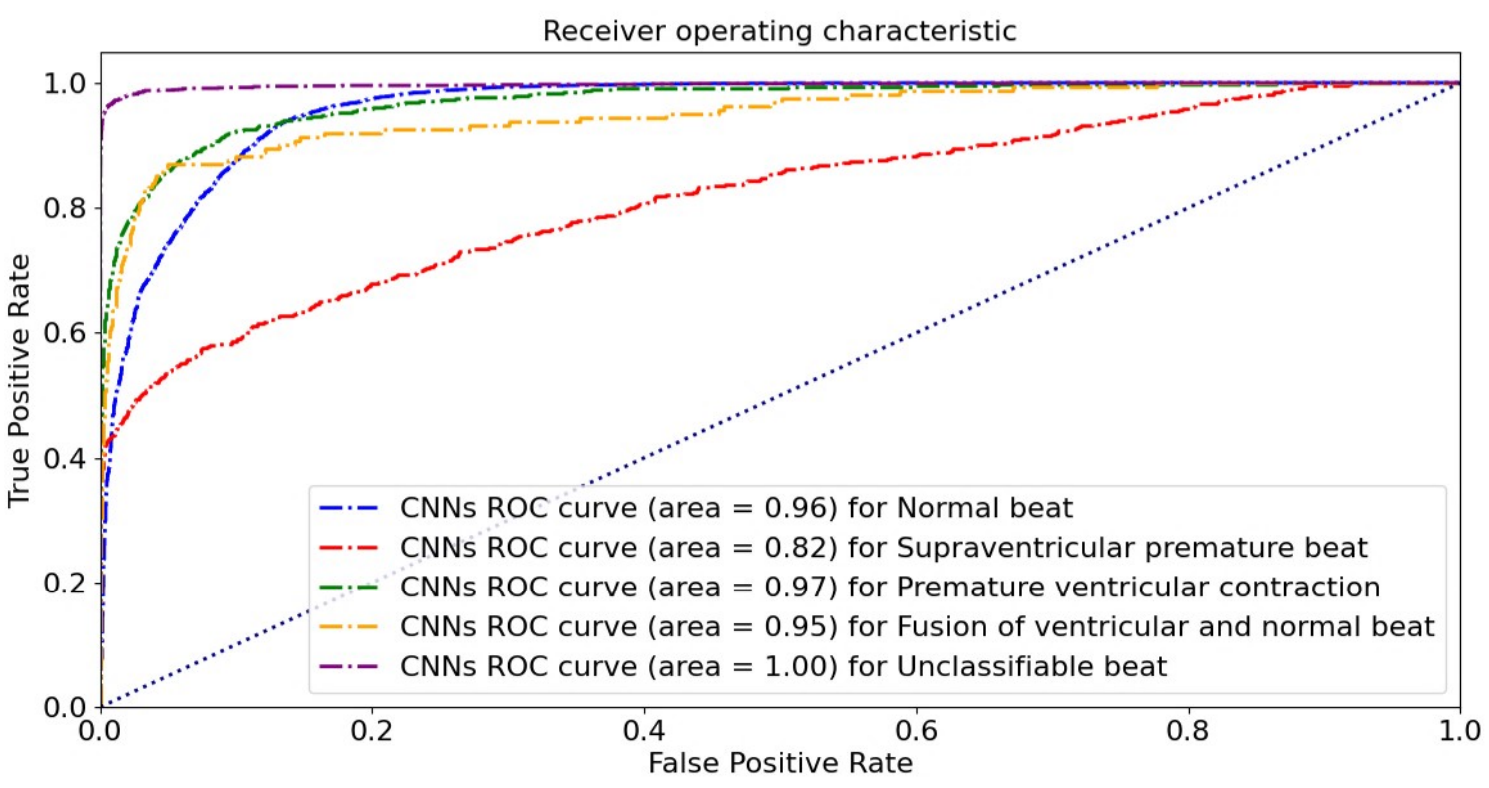}
\includegraphics[width=0.435\linewidth]{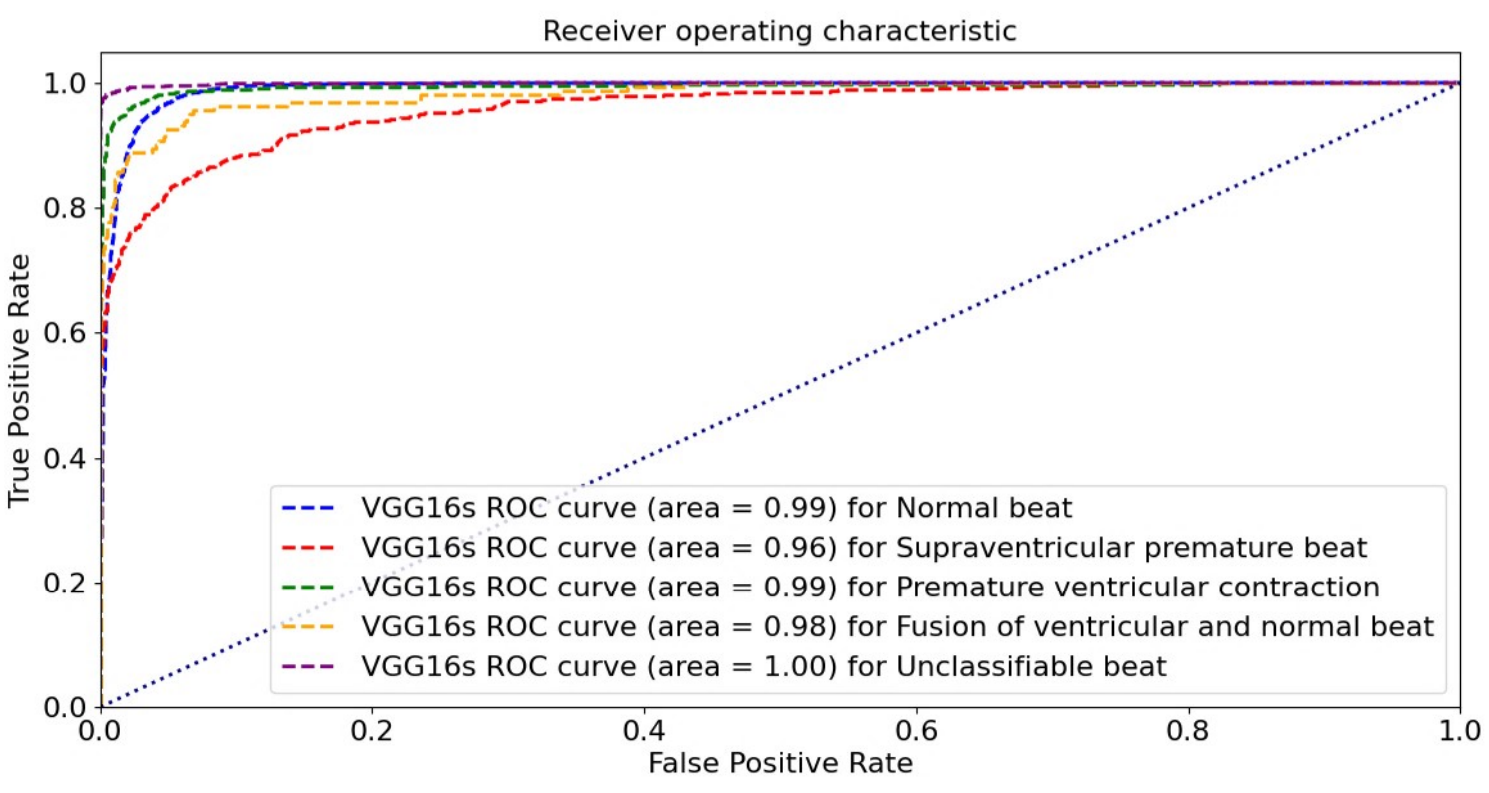}\\
(a)~~~~~~~~~~~~~~~~~~~~~~~~~~~~~~~~~~~~~~~~~~~~~~~~~~~~~~~~~~~~~~~~~~~~~~~(b)\\
\includegraphics[width=0.435\linewidth]{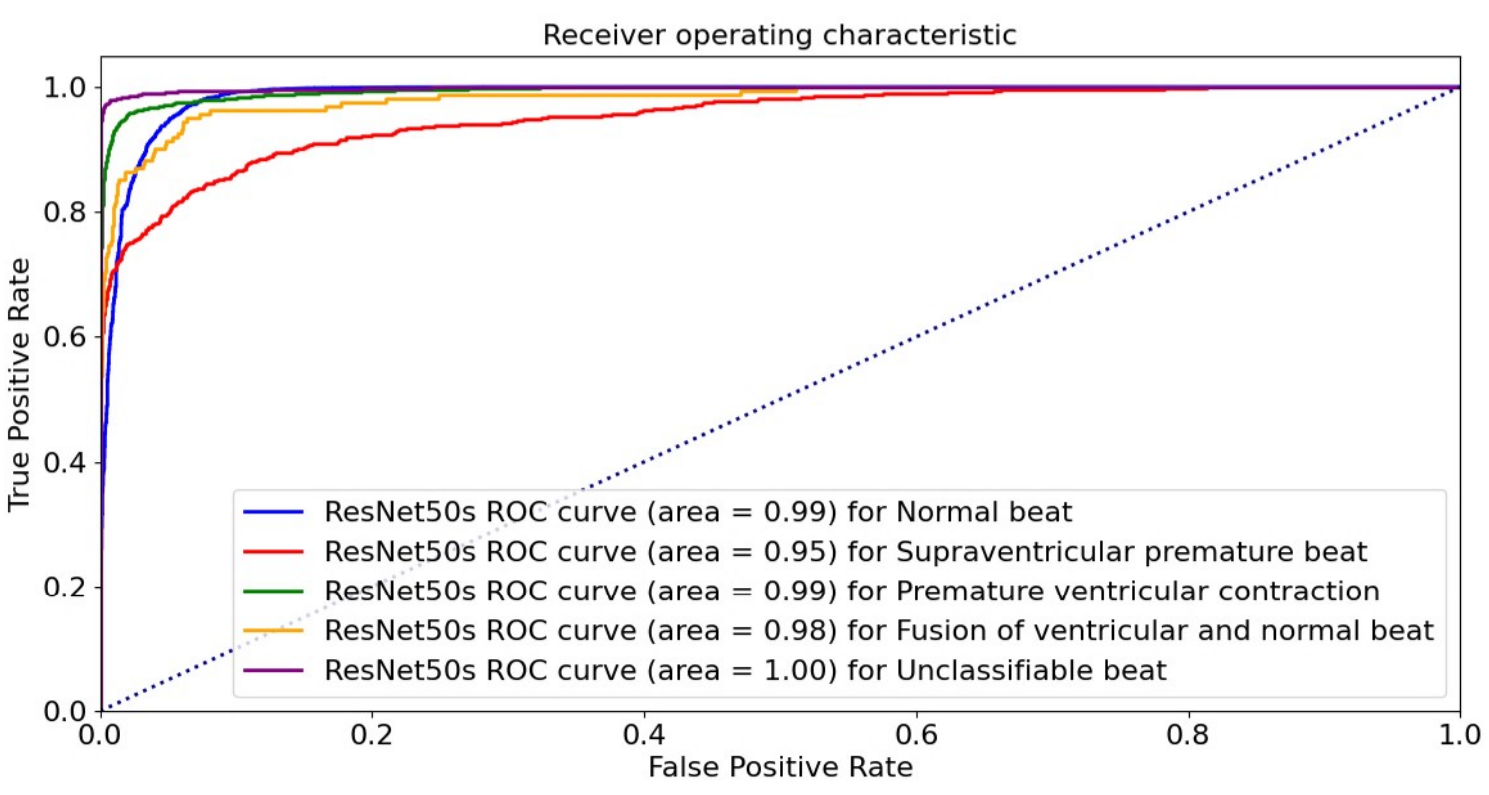}
\includegraphics[width=0.435\linewidth]{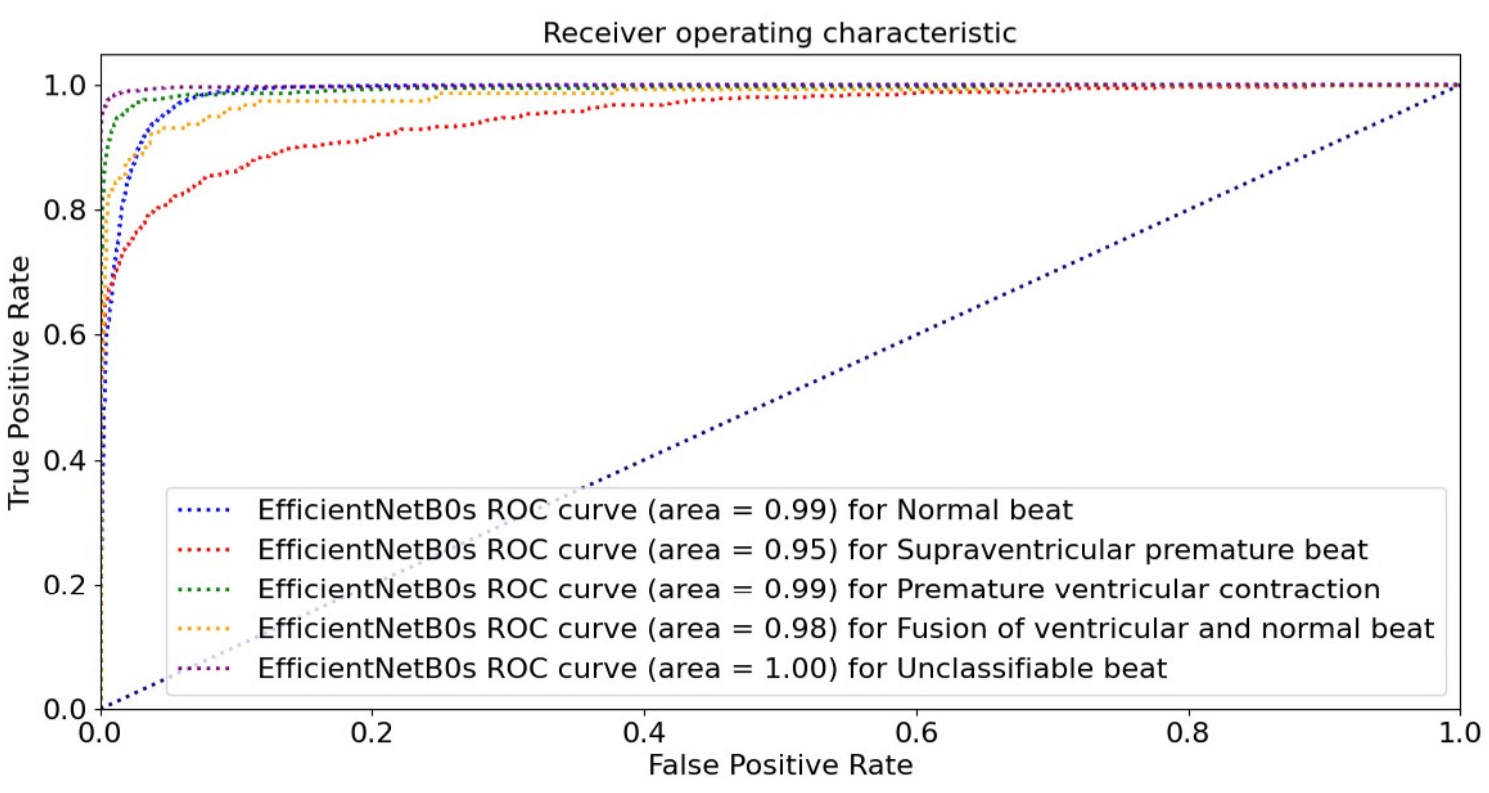}
(c)~~~~~~~~~~~~~~~~~~~~~~~~~~~~~~~~~~~~~~~~~~~~~~~~~~~~~~~~~~~~~~~~~~~~~~~(d)\\
\includegraphics[width=0.6\linewidth]{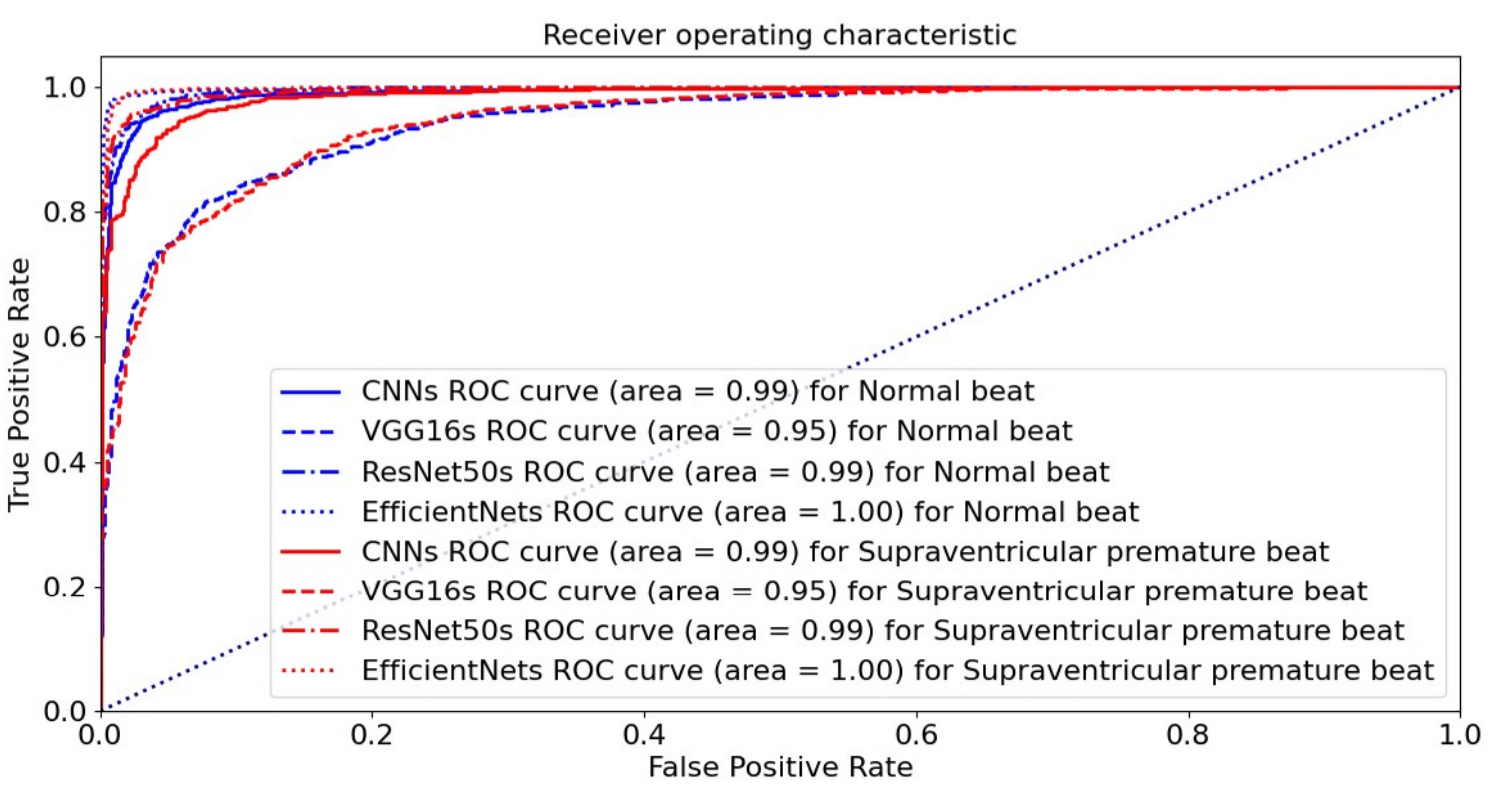}
\\(e) The PTB Diagnostic \\
\caption{ROC curves of Arrhythmia using (a) CNN, (b) VGG-16, (c) ResNet-50 and (d) EfficientNet.}
\label{fig:ROC}
\end{figure*}

\begin{figure*}[!t]
\centering
\includegraphics[width=0.395\linewidth]{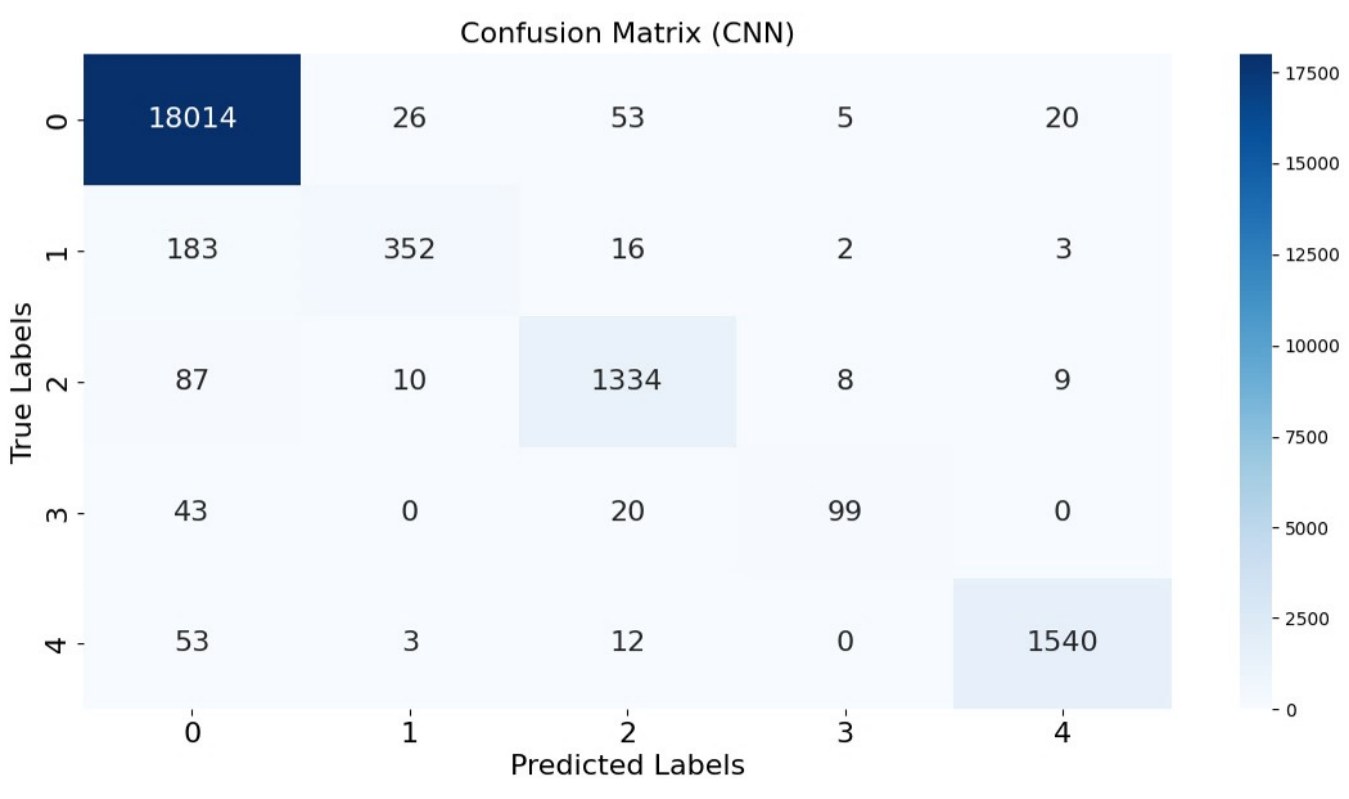}
\includegraphics[width=0.395\linewidth]{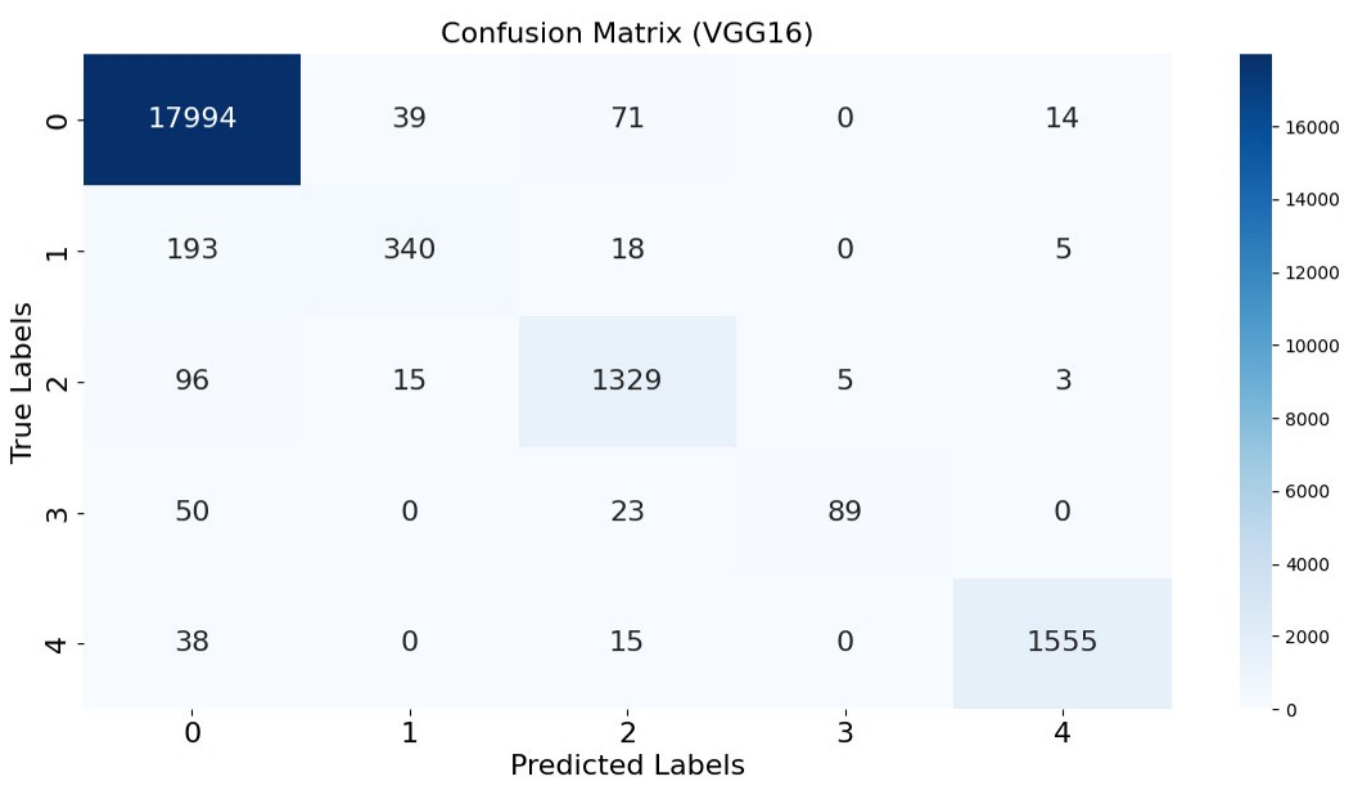}\\
(a)~~~~~~~~~~~~~~~~~~~~~~~~~~~~~~~~~~~~~~~~~~~~~~~~~~~~~~~~~~~~~~~~~~~~~~~(b)\\
\includegraphics[width=0.395\linewidth]{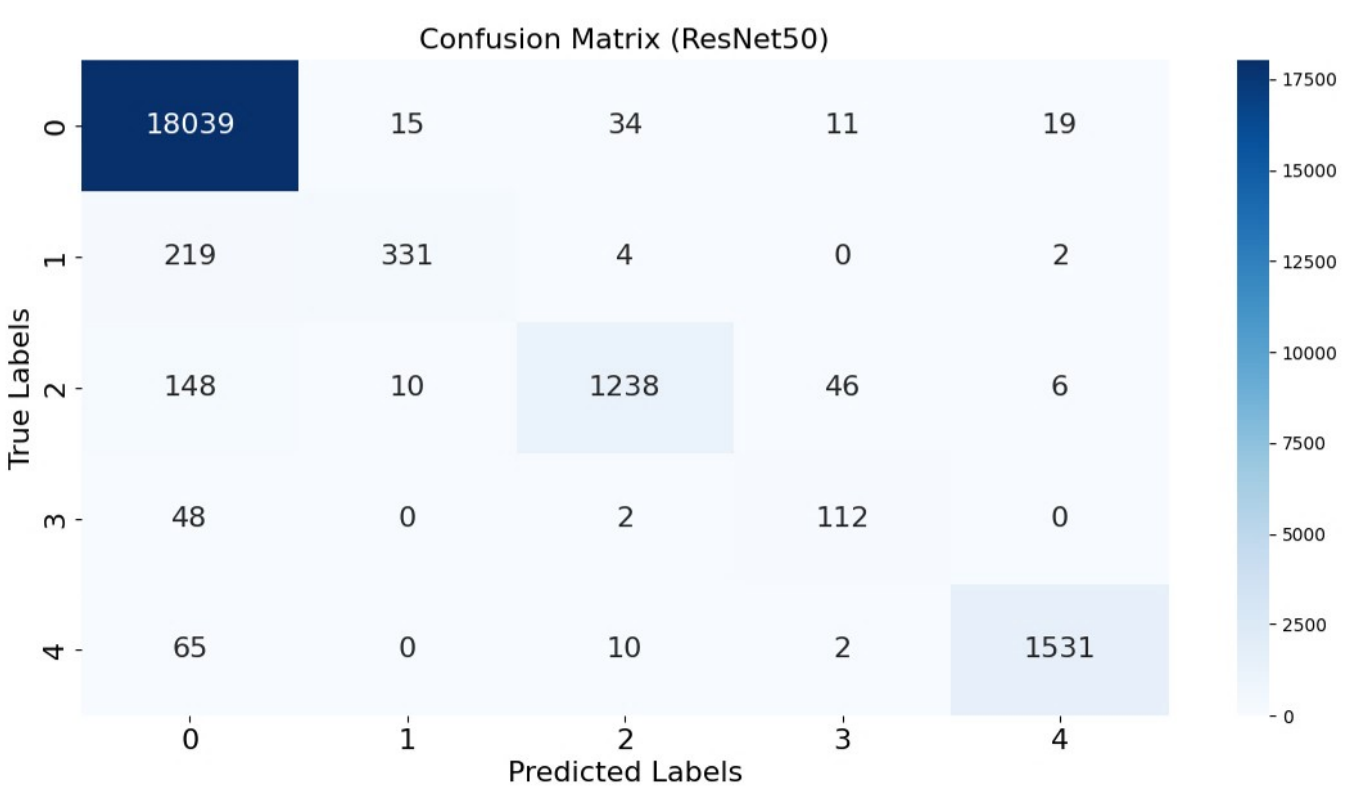}
\includegraphics[width=0.395\linewidth]{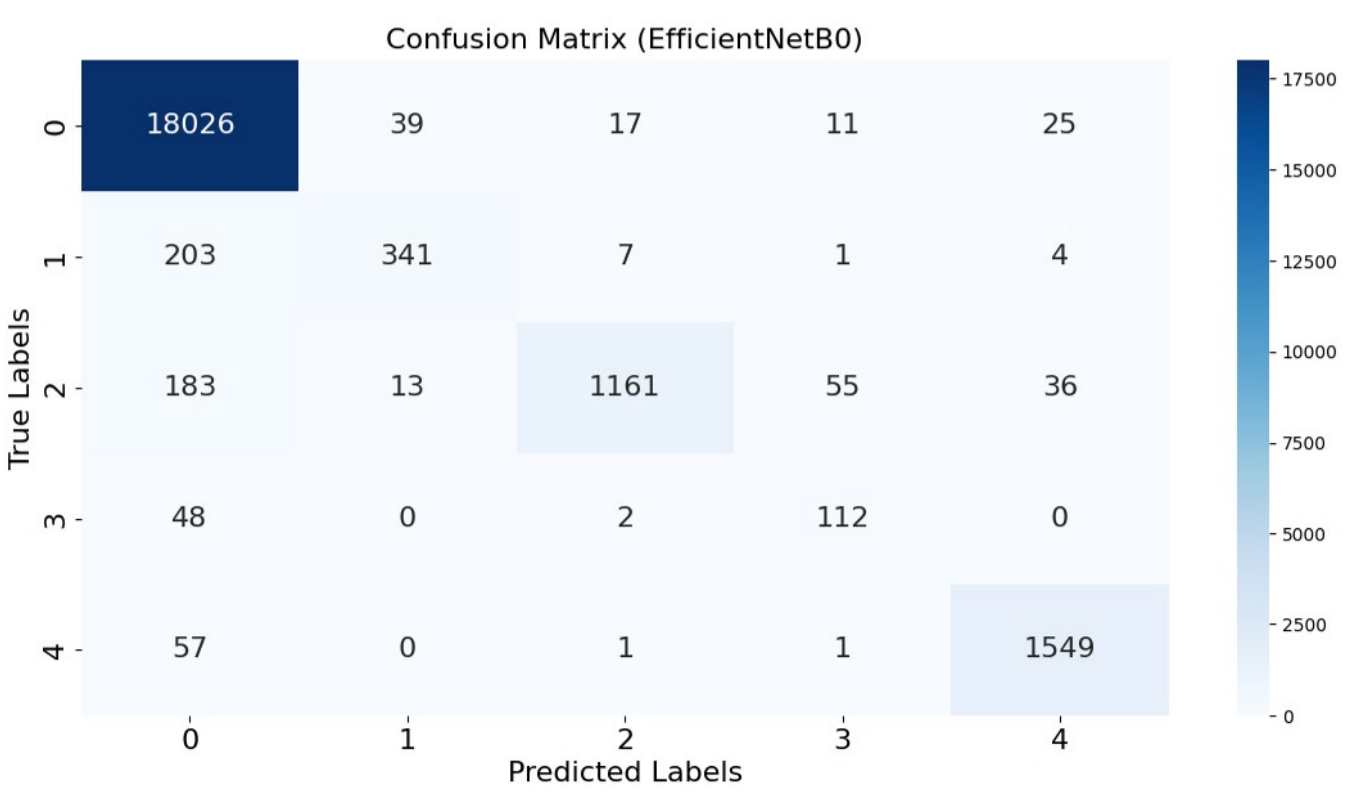}
\caption{Confusion matrices under the Arrhythmia dataset using (a) CNN, (b) VGG-16, (c) ResNet-50 and (d) EfficientNet.}
\label{fig:Confusion matrix}
\end{figure*}

The best obtained classification accuracy is 97.47\%. All other results are presented in details in Table \ref{tab1}, Figures \ref{fig:ROC} and \ref{fig:Confusion matrix}.\\

\subsection{Discussion}
Considering the differences in dataset size and class distribution, the results achieved with various methods, including the use of GAF as a common feature representation technique, for ECG classification on the Arrhythmia and PTB Diagnostic datasets are noteworthy. The Arrhythmia dataset, with its larger sample size of 109,446 samples and 5 classes, demonstrates impressive accuracy and F1-score across multiple methods, including CNN (97.47\% accuracy, 97.34\% F1-score), VGG16 (97.32\% accuracy, 97.18\% F1-score), ResNet50 (97.07\% accuracy, 96.92\% F1-score), and EfficientNet (96.78\% accuracy, 96.64\% F1-score), all using GAF as a feature representation technique.

Similarly, the PTB Diagnostic dataset, with its smaller sample size of 14,552 samples and only 2 classes, also achieves high accuracy and F1-score with the different methods, including CNN (97.56\% accuracy, 97.56\% F1-score), VGG16 (89.17\% accuracy, 88.98\% F1-score), ResNet50 (95.84\% accuracy, 95.90\% F1-score), and EfficientNet (98.65\% accuracy, 98.65\% F1-score), all utilizing GAF as a common feature representation technique.

These results indicate that CNN, VGG16, ResNet50, and EfficientNet methods, when used in conjunction with GAF as a feature representation technique, are effective in ECG classification tasks. The performance of these methods on both Arrhythmia and PTB Diagnostic datasets, with the utilization of GAF, highlights the robustness and generalizability of these methods for ECG classification tasks.

\label{sec:results}

\section{CONCLUSION}
\label{sec:conclusion}
 The experiments using Gramian Angular Field (GAF) with CNN for ECG classification yielded promising results, demonstrating high accuracy and F1-score on both the Arrhythmia and PTB Diagnostic datasets. These findings suggest that GAF is a highly effective method for ECG classification, regardless of dataset size and class complexity.
Although the use of pretrained models did not result in significant improvement in the current experiments, it is important to note that fine-tuning them may require further optimization and resources to fully leverage their potential for ECG classification. Future research should investigate different pretrained models and their applicability to ECG classification.

To further improve ECG classification, future research could explore comparing GAF-based classification with other approaches, such as time-frequency based methods, or fusing time-frequency features with image features extracted from GAF. This could provide insights into the comparative performance and potential synergies of different techniques, contributing to the development of more accurate and robust classification models in the future.
In conclusion, the results of our study suggest that GAF has significant potential for improving ECG signal classification and warrants further investigation. Further research and optimization of pretrained models, as well as exploring different approaches and techniques, could advance ECG classification methods for clinical applications.


\end{document}